# Granular permittivity representation in extremely near-field light-matter interactions processes

Alexey S. Kadochkin[1,2,*], Alexander S. Shalin[2], and Pavel Ginzburg[2,3]

[1] Ulyanovsk State University, Ulyanovsk, 432017, Russia
[2] "Nanooptomechanics" Laboratory, ITMO University, St. Petersburg, 197101, Russia
[3] School of Electrical Engineering, Tel Aviv University, Tel Aviv, 6997801, Israel

**Abstract:**

Light-matter interaction processes are significantly affected by surrounding electromagnetic environment. Dielectric materials are usually introduced into an interaction picture via their classical properties, e.g. permittivity, appearing in constitutive relations. While this approach was proven to be applicable in many occasions, it might face limitations when an emitter is situated very close to a material boundary. In this case nonlocal extend of a quantum wave function of an emitter becomes comparable with a distance to a boundary and a lattice constant of a material. Here a semi-classical model, taking into account material's granularity, is developed. In particular, spontaneous emission process in the vicinity of flat boundaries is considered. The material boundary is divided into a pair areas - far zone is modeled as a continuous phase, while the near zone next to a nonlocal emitter is represented with a discrete array of polarizable particles. This array resembles optical properties of the continuous phase under the standard homogenization procedure. Local field effects were shown to lead orders of magnitude corrections to spontaneous emission rates in the case of sub-nanometer emitter-surface separation distances. The developed mesoscopic model enables addressing few aspects of local field corrections in quite complex scenarios, where quantum *ab initio* techniques yet face challenges owing to involved computational complexity. The developed method could be utilized for designs of quantum sources and networks, enhanced with structured electromagnetic environment.

\* Corresponding author: askadochkin@sv.ulsu.ru

**Introduction**

Flexible control over light-matter interaction dynamics with structured media enables demonstration of many fundamental effects and powers numerous practical applications [1]. The key concept, capable of describing majority of interaction processes, relies on introduction of photonic density of states (DOS) [2]. While this quantity in free space is mostly defined by an operation frequency, DOS could be locally manipulated by introducing various electromagnetic structures. In this case, local density of states (LDOS) tailors majority of light-matter interaction processes, such as spontaneous emission on the most fundamental level [2]. Acceleration of spontaneous emission rates in structured environment respectively to a free space is called Purcell effect [3] and its value is directly proportional to LDOS. While traditional pathways for accelerating spontaneous emission and even reaching strong coupling regimes of interaction utilize high quality factor cavities, recent advances in nanofabrication offer complementary solutions of using small subwavelength resonators [4]. This approach of open type of cavities suggests manipulation of light-matter interaction processes via local field control or, in other words, via small modal volumes [5], [6], [7], [8], [9]. For example, metallic antennas, supporting localized plasmonic resonances, were shown to provide flexible solution for achieving moderate Purcell enhancement and directionality in emission, e.g. [5], [10]. In contrary to traditional cavities case, the near-field coupling approach requires placing emitters in a close proximity to material boundaries. Treatment of separation distances, smaller than 10 nm, requires extra-care, since quenching mechanisms start playing a role, e.g. [11], [12], [13]. Furthermore, those extremely short range scenarious might question classical approaches, utilized for description of light-matter interaction processes in the presence of material bodies [14]. Recent experimental studies demonstrate interactions between molecules and pico-cavities (one or few missing atoms in a metal surface) [15]. In general, *ab initio* methods, such as density functional theory, or other semi-phenomenological quantum approaches (e.g. [16] for a review) are required for description of those extreme scenarios. However, a comprehensive modeling of a structure 'atom by atom' is still very involved due to an enormous computation complexity, required for calculations. Nevertheless, classical models, supplemented with a certain number of fitting parameters, are still found successful in description of short range interactions with material boundaries [17]. Consequently, further development of mesoscopic models is very important in order to cope with a large span of processes, where closely situated material boundaries are involved.

Here a mesoscopic model of light-matter interactions next to a material interface is developed and the process of spontaneous emission is analyzed in details. While majority of studies on this topic utilizes a point dipole description of an emitter (e.g. [18], [19]), quantum wavefunctions of the later could have remarkable sizes. For example, polaritons at cryogenic temperatures could have wavefunctions, coherently spreading along several microns. Majority of organic dyes, widely used as florescent tags, are also not point dipoles, but to a much smaller extend - nanometers and less. However, this scale becomes important when those dye molecules are brought into proximity to dielectric materials that have comparable lattice constants. In this case, a discrete nature of the material lattice should be taken into account. The mesoscopic model, developed here, comes to address the impact of material granularity on emission processes from nonlocal (not a point dipole) emitters. In particular, the continuous material phase is represented by an array of polarizable deeply subwavelength spheres, converging to the classical permittivity response in a macroscopic average (Fig. 1).

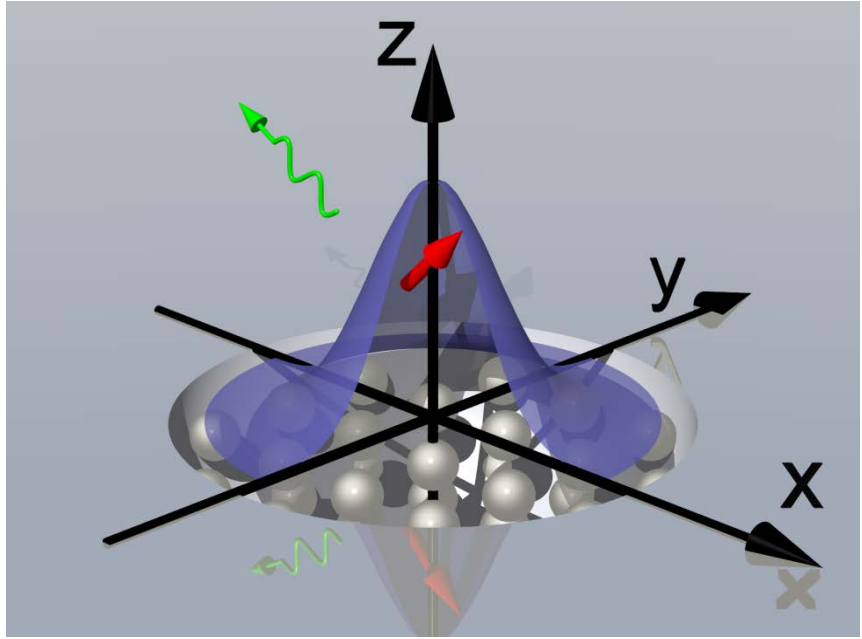

Fig. 1. Illustrative representation of spontaneous emission next to a flat surface. Quantum emitter is represented by a spatially distributed wavefunction (blue Gaussian). The material underneath the emitter is separated to a granular near zone (array of grey spheres) and a continuous phase (far zone), described by a classical permittivity function.

Discretized representation of a material boundary was compared to standard continuous models and deviations of orders of magnitudes in spontaneous emission rates were observed in the case of nanometric separation distances between an emitter and the structure.

**Results and Discussion**

*Quantum theory of emission from nonlocal objects in material environment*

Quantization of electromagnetic field in a presence of material bodies, having nontrivial dispersion and absorption properties, requires extra care. Many theoretical approaches have been developed in order to account for material degrees of freedom, but resulting Hamiltonians are hardly diagonalizable if nontrivial geometries of material bodies are involved [20]. A widely employed approach, capable to treat light-matter interactions in a presence of complex electromagnetic structures, utilizes Langevin quantization [14]. The resulting Hamiltonians depend on classical electromagnetic Green's functions, and macroscopic material susceptibilities are introduced phenomenologically (bulk properties are used in this case). One especially important problem, to be addressed once applying this method, is the proper consideration of the embedding environment. For example, imaginary part of the Green function (proportional to LDOS) will diverge in a lossy medium. Several local field correction approaches were developed [21], [22], [23] and they suggest excluding a void (depolarization) volume from a bulk. However, the shape and size of this depolarization volume strongly affects the resulting spontaneous emission rate, motivating development of other approaches. The model, developed hereafter, enables relaxing constrains of classical bulky materials representation, and account for finite size of the lattice constant. In this case, the general Langevin quantization procedure still could be employed and the spontaneous emission rate from nonlocal (not point-like) is given by kernel integration as [24]:

$$\gamma = \frac{2\omega_0^2}{\varepsilon_0 \hbar c^2} \text{Im}\left\{ \iint d^3\vec{r} d^3\vec{r}\,' \vec{\mu}(\vec{r}) \vec{G}(\vec{r},\vec{r}\,',\omega_0) \vec{\mu}(\vec{r}\,') \right\}, \quad (1)$$

where $\omega_0$ is the carrier frequency of the emitted photon, $\vec{G}(\vec{r},\vec{r}\,',\omega_0)$ is the classical electromagnetic Green's function, and $\vec{\mu}(\vec{r})$ is the dipolar density of the emitter. In the local case, when the spatial distribution of the electron wavefunction could be neglected, the dipolar moment resembles its widely acceptable form of $\vec{\mu} = e\int \psi_e^*(\vec{r}) \vec{r} \psi_g(\vec{r}) d^3\vec{r}$ with $\psi_e$ and $\psi_g$ being excited and ground wavefunctions respectively, while $e$ is electron charge. Numerical investigations hereafter rely on Eq. 1 and utilize classical Gaussian distribution of a currents, similar to approaches, developed at [24]. Those results will be compared with a classical point-dipole approach, which completely neglects a possible nonlocality of an emitter. Parameters of Gaussian distributions of dipolar densities, which appear at Eq. 1, correspond to typical parameters of Rhodamine family of fluorescent dyes. Typical sizes of those molecules, obtained from hydrodynamic measurements, are around 1.2 [25], which is larger than lattice constants of dielectric substrates (typically 0.5 nm).

*Combined discrete-continuous representation of material bodies*

Macroscopic Maxwell's equations are derived from the microscopic set with an appropriate homogenization of material responses [26]. This procedure leads to reliable results, if k-vector of an incident radiation is much smaller than a k-vector of a material lattice (in other words, k-vector of a wave does not approach edges of Brillouin zone of a homogenized material). In majority of the cases those conditions are satisfied and local material parameters could be introduced within the constitutive relations. However, if a wave accumulates a considerable phase along it's propagation within a unit cell, effects of spatial dispersion start playing a role and local material parameters become k-dependent [27]. Another case, when this effect could be important, is related to a scenario, where localized quantum emitter situated close or within a material body. In this case the plane wave representation of the source should include a large span of k-vectors and a part of their spectrum could touch edges of material's Brillion zone. In this case, spatial dispersion effects should be also introduced. This procedure could be relatively straightforward, if a localized emitter is situated next to a planar surface – in this case, knowledge of Fresnel coefficients is sufficient for appropriate calculations, e.g. [18]. However, this procedure will become extremely complicated, if nontrivial geometries are involved. Materials' representation, as a set of polarazable point dipoles, placed in crystal nodes, goes back to Hopfield's model [28] and was recently applied for resolving divergent emission rates in hyperbolic metamaterials [29]. However, special lattice summation techniques should be employed and they are not applicable in the case of finite, yet large, geometries. The major goal of this work is to develop semi-discrete phenomenological model, which treats a close proximity of an emitter as a discrete array of small polarizable spheres, while the far zone is introduced as a continuous classical bulk material (Fig. 1). This approach enables phenomenological treatment of arbitrary geometries by separating relevant areas to near and far zone representation.

Array of polarizable spheres, to be introduced in the near zone, was homogenized via standard Lorentz-Lorenz approach. Two types of materials will be considered hereafter – transparent dielectric with refractive index ($n$) of 1.5 (glass) and a negative permittivity metal with $n = 0.052277 + 3.9249 \cdot i$, corresponding to silver at 590nm wavelength (central emission line of Rhodamine 6G, as an example). In the following phenomenological qualitative analysis, the crystal lattice was taken to be primitive cubic with a lattice constant of $a = 0.5$ nm. Furthermore, strictly speaking, nonlocality in noble metals originates from collective dynamics of conduction electrons, e.g. [30], [31]. Therefore, discrete spheres array model in this case should be considered as a phenomenological approach only and without a relation to solid state aspects.

*Spontaneous emission next to a glass surface*

The first scenario, which will be analysed, is the spontaneous emission next to a semi-infinite dielectric surface. Refractive index of a bulk glass was taken to be $n = 1.5$, while the discrete array was comprised of spheres with radiuses $R_p = 0.2$ nm, separated by 0.5 nm distances. The dielectric constant $\varepsilon_p$ was chosen in the way, that the individual particle's polarizability $\alpha_p = 4\pi\varepsilon_0 R_p^3 (\varepsilon_p - 1)/(\varepsilon_p + 2)$ will satisfy Lorentz-Lorenz condition for average refractive index - $(n^2 - 1)/(n^2 + 2) = N\alpha_p/(3\varepsilon_0)$ [2]. The same approach will be used for the metal case hereafter. The number of spheres, sufficient for representing the discrete phase of matter was obtained empirically. The size of the arrays was continuously increased until a convergence of Purcell factor was observed [32]. Five spheres at the top level were found to be sufficient (Fig. 5 for schematics).

Fig. 2 shows values of Purcell factors for all four scenarios – two types of emitters ((i) Point dipole (ii) Gaussian distribution of dipolar density) and two types of material representations ((i) continuous medium and (ii) combined discrete-continuous representation).

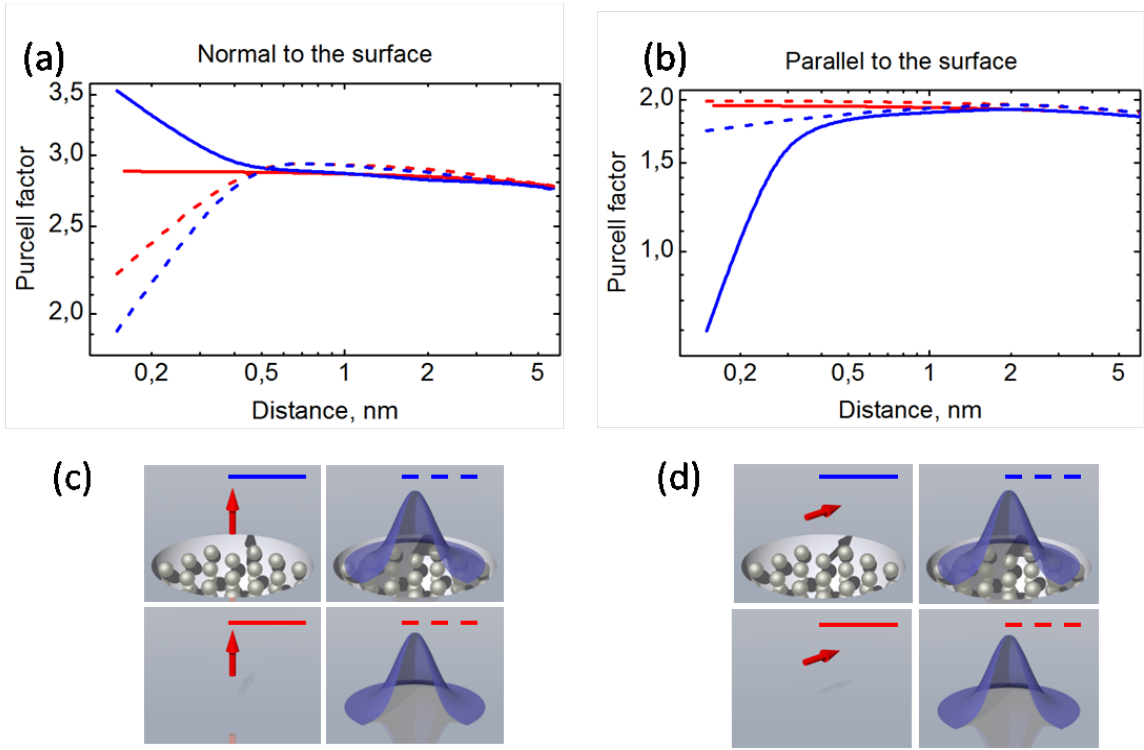

Fig. 2. Purcell factor near semi-infinite glass surface (glass, *n*=1.5). Different lines correspond to four different scenarios, indicated in the insets below the panels and elaborated in the main text. (a) Emitter, parallel to the surface; (b) Emitter, perpendicular to the surface.

First, one can see that Purcell enhancements converge to the same value, once the distance between the emitter and the surface become several nanometers (few lattice constants). At the large distance limit the values correspond well to classical analytical formulas for local point emitters [2]. However, significant deviations start emerging at distances smaller than 1 nm. In the case of the emitter, perpendicular to the surface, Gaussian model shows about 30% lower Purcell enhancement and the smallest value is obtained for the combined discrete-continuous surface representation. In the case of the emitter, parallel to the surface, the smallest value is demonstrated by the point dipole above the combined model for the surface. It is worth noting, that about 50% quenching is observed in this case, while other models suggest modest Purcell enhancement.

*Spontaneous emission next to a silver surface*

Emission processes next to metallic structures are qualitatively different from processes, where transparent dielectric components are involved. Noble metals, such as silver here, have negative permittivity at the visible and infrared spectral ranges. Structures, made of noble metals, could support localized plasmon resonances, which are very beneficial e.g. for fluorescence enhancement. On the other hand, inherent material losses of metal components lead to nonradiative quenching. The competition between radiative and nonradiative enhancement is a key for understanding those emission processes. In this section the same set of phenomenological models, used for describing emission next to the glass surface, will be applied.

Three parameters, characterizing the emission, will be considered. The total enhancement stands for the overall acceleration of recombination, while the radiative enhancement corresponds to the far-field contribution. The first quantity was calculated by integrating the Poynting vector flux via a small sphere, encapsulating the emitter. The radiative contribution was evaluated with the similar integration at the far-field zone. Fluorescent quantum yield was calculated as the ratio between total and radiative enhancements, while internal quantum yield of the emitter was taken to be unity. Similarly to the case of the transparent dielectric substrate, all four models converge to the same values, once the separation distances are several nm and more (Fig. 3 and 4). At shorter distances, however, significant deviations are observed. The granular representation for two types of emitters suggests higher total enhancement in both orientations of the dipolar moment (Fig. 3 for the dipole, perpendicular to the surface, Fig. 4 for the parallel case). Remarkably, the radiative enhancement is also higher for the granular representation. However, the true measurable quantity in majority of experiments (e.g. time correlated single photon counting) is the fluorescent quantum yield. The quantum yield in all of the cases drops to zero and replicates the well-known effect of nonradiative quenching, where the energy goes into material excitations and dissipates. However, the discrete material representation predicts much more dramatic drop of the quantum yield (almost two orders of magnitude difference) in respect to homogeneous permittivity models.

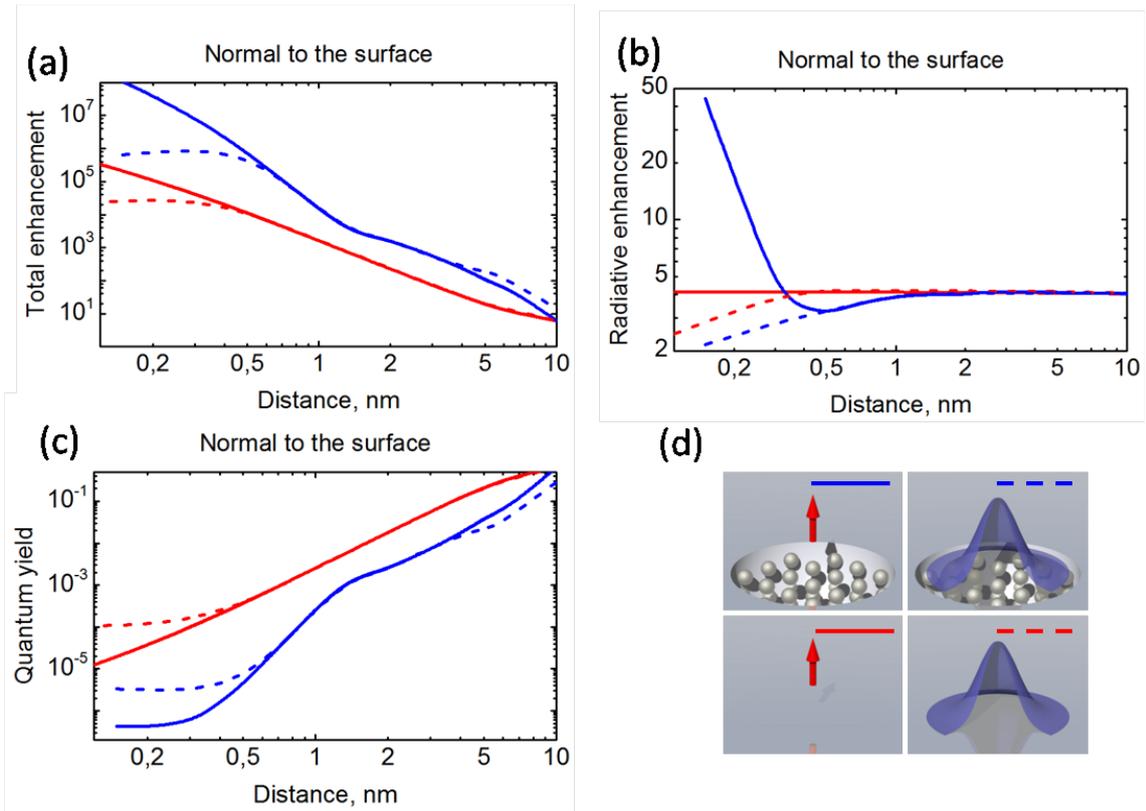

Fig. 3. Emission near semi-infinite silver surface (emission wavelength 590nm, emitter's dipole moment is perpendicular to the surface). Lines correspond to four different scenarios, indicated in the insets (panel (d)). (a) Total rate enhancement. (b) Radiative enhancement. (c) Quantum yield.

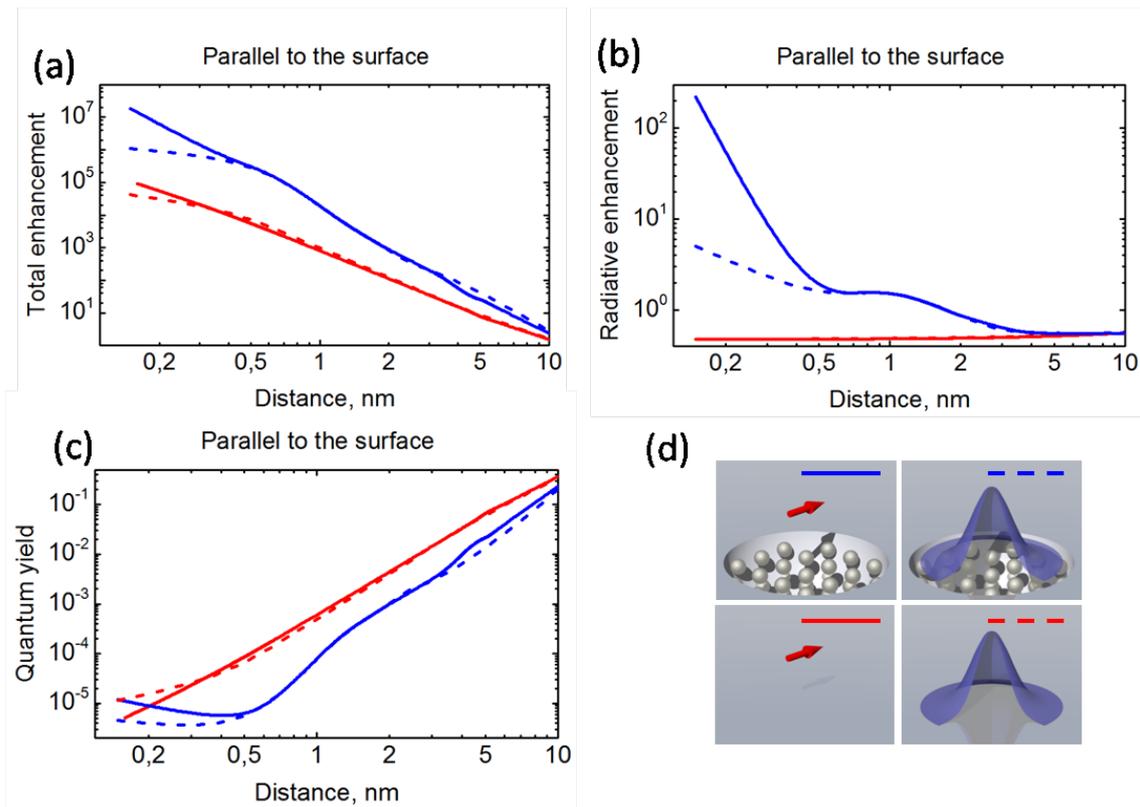

Fig. 4. Emission near semi-infinite silver surface (emission wavelength 590nm, emitter dipole moment is parallel to the surface). Lines correspond to four different scenarios, indicated in the insets (panel (d)). (a) Total rate enhancement. (b) Radiative enhancement. (c) Quantum yield.

Differences between the models can be also identified by observing near-field distributions nearby the emitters. Fig. 5 shows field amplitudes (logarithmic scale) in the case of 0.1 nm separation distance between the emitter (center of distribution for Gaussian source) and the surface. The comparison is made for the granular representation models of glass and silver surfaces and for the Gaussian and point dipole emitters, oriented perpendicularly to the surface (this is the most representative case). It can be seen that the point source in both cases exhibits a local maxima around itself, while the distributed emitter has a dark spot. This behavior is much more pronounced for the silver substrate. Since the total Purcell enhancement is proportional to the imaginary part of the Green function, the point dipole demonstrates larger overall rates in the full correspondence with the results, demonstrated on Fig. 3(a). Gaussian distribution model enable smearing additional possible singularities (for both dipole moment orientations) and, as the result, leads to less extreme values of both total and radiative enhancements (Fig 3). Furthermore, exponentially decaying tails of the Gaussian distribution overlaps with the medium, which leads to additional quenching. This simplistic model could also be phenomenologically related to effects of electron transfer and tunnelling to the surface.

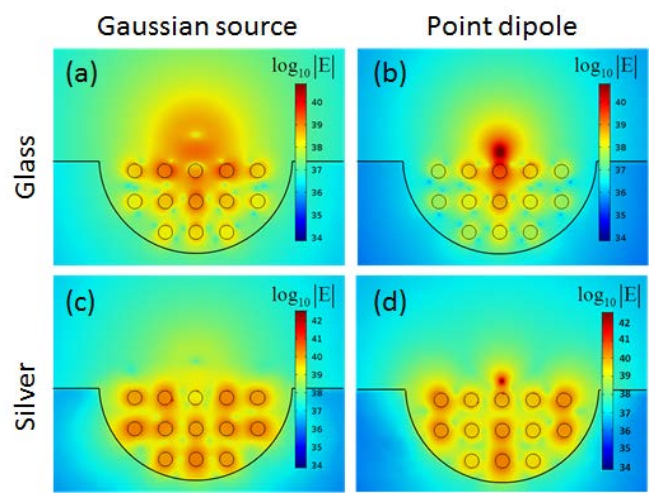

Fig. 5. Electric field (in arbitrary units) amplitude distribution (emission wavelength 590nm, dipole moment is perpendicular to the surface)(a), (b) – glass substrate, (c), (d) – silver substrate, (a),(c) – Gaussian emitter, (b), (d) – point dipole emitter.

**Summary and Conclusions**

Light-matter interaction dynamics, affected by nearby electromagnetic structures has been considered. In particular, a regime, where a quantum emitter is situated at sub-nanometric distance from a surface, was analyzed phenomenologically. A set of mesoscopic models ((i) Point dipole (ii) Gaussian distribution of dipolar density) and two types of material representations ((i) continuous medium and (ii) combined discrete-continuous representation)) was compared with a semi-classical description, which utilizes a point dipole next to a homogeneous epsilon bulk description. While the semi-classical approach was found to be sufficient in the cases when the separation distances lager than several nanometers, it failed to describe scenarios where relevant gaps become comparable with the lattice constants of material components.

It was shown that Gaussian model for the emitter and granular representation of the matter models has advantages in describing quenching phenomena. It is also worth noting that nearby dielectric bodies could significantly affect quantum structure of an emitter, e.g. leading to level shifts etc. In this case, the proposed mesoscopic model will require additional modifications in order to approach a self-consistent description of interaction processes.

**Acknowledgements:**

P.G. acknowledges the support from TAU Rector Grant, PAZY foundation, and German-Israeli Foundation (GIF, grant2399). The calculations of field distributions have been supported by the Russian Science Foundation Grant No. 16-12-10287; A.S. acknowledges the support of the President of Russian Federation in the frame of Scholarship SP-4248.2016.1, the support of Ministry of Education and Science of the Russian Federation in the frame of GOSZADANIE Grant No. 3.4982.2017/6.7 and partial support of Russian Foundation for Basic Research (16-52-00112). A.K. acknowledges the support of Ministry of Education and Science of the Russian Federation (Project No. 14.Z50.31.0015)